\documentclass{agujournal2018}
\usepackage{apacite}
\usepackage{url} 

\usepackage{hyperref}
\usepackage{makecell}
\usepackage{arydshln, collcell}
\usepackage{subcaption}
\usepackage{colortbl}
\usepackage{array}
\usepackage{textcomp}
\usepackage[flushleft]{threeparttable}

\draftfalse

\newcommand{\aap}{    {\it Astron. Astrophys. }}

\newcommand{\apjl}{   {\it Astrophys. J. Lett. }}
\newcommand{\apjs}{   {\it Astrophys. J. Suppl. S.}}

\graphicspath{{./}{figures/}}

\journalname{Space Weather}

\begin{document}

%
%


\title{Why are ELEvoHI CME arrival predictions different if based on STEREO-A or STEREO-B heliospheric imager observations?}

%
%




\authors{J\"urgen Hinterreiter\affil{1,2}, Tanja Amerstorfer\affil{1}, Martin A. Reiss\affil{1,3}, Christian M\"ostl\affil{1,3}, Manuela Temmer\affil{2}, Maike Bauer\affil{1,2}, Ute V. Amerstorfer\affil{1}, Rachel L. Bailey\affil{4}, Andreas J. Weiss\affil{1,2,3}, Jackie A. Davies\affil{5}, Luke A. Barnard\affil{6}, Mathew J. Owens\affil{6}}
\affiliation{1}{Space Research Institute, Austrian Academy of Sciences, Schmiedlstraße 6, 8042 Graz, Austria}
\affiliation{2}{University of Graz, Institute of Physics, Universitätsplatz 5, 8010 Graz, Austria}
\affiliation{3}{Institute of Geodesy, Graz University of Technology, Steyrergasse 30, 8010 Graz, Austria}
\affiliation{4}{Conrad Observatory, Zentralanstalt f\"ur Meteorologie und Geodynamik, Vienna, Austria}
\affiliation{5}{RAL Space, Rutherford Appleton Laboratory, Harwell Campus, Didcot OX11 0QX, UK}
\affiliation{6}{Department of Meteorology, University of Reading, Reading, UK}





\correspondingauthor{J\"urgen Hinterreiter}{juergen.hinterreiter@oeaw.ac.at}




\begin{keypoints}
\item A comparison of CME arrival time and speed predictions from two vantage points was carried out using ELEvoHI
\item A highly structured ambient solar wind flow leads to larger arrival time differences between STA and STB predictions
\item The assumption of a rigid CME front in ELEvoHI and other HI-based methods is most probably too simplistic
\end{keypoints}

%
%


\begin{abstract}
Accurate forecasting of the arrival time and arrival speed of coronal mass ejections (CMEs) is a unsolved problem in space weather research. In this study, a comparison of the predicted arrival times and speeds for each CME based, independently, on the inputs from the two STEREO vantage points is carried out. We perform hindcasts using ELlipse Evolution model based on Heliospheric Imager observations (ELEvoHI) ensemble modelling. An estimate of the ambient solar wind conditions is obtained by the Wang-Sheeley-Arge/Heliospheric Upwind eXtrapolation (WSA/HUX) model combination that serves as input to ELEvoHI. We carefully select 12~CMEs between February 2010 and July 2012 that show clear signatures in both STEREO-A and STEREO-B HI time-elongation maps, that propagate close to the ecliptic plane, and that have corresponding in situ signatures at Earth. We find a mean arrival time difference of 6.5~hrs between predictions from the two different viewpoints, which can reach up to 9.5~hrs for individual CMEs, while the mean arrival speed difference is 63~km~s$^{-1}$. An ambient solar wind with a large speed variance leads to larger differences in the STEREO-A and STEREO-B CME arrival time predictions ($cc~=~0.92$). Additionally, we compare the predicted arrivals, from both spacecraft, to the actual in situ arrivals at Earth and find a mean absolute error of 7.5~$\pm$~9.5~hrs for the arrival time and 87~$\pm$~111~km~s$^{-1}$ for the arrival speed. There is no tendency for one spacecraft to provide more accurate arrival predictions than the other.

\end{abstract}

%
%

%


%
%
%
%

\section{Introduction} \label{sec:intro}

Understanding the dynamics of coronal mass ejections (CMEs) in the heliosphere is a key aspect of space weather research. CMEs are huge clouds of energetic and magnetized plasma \citep{Hundhausen1994} erupting from the solar corona that may reach speeds of up to 3000~km~s$^{-1}$. When they hit Earth, CMEs can produce strong geomagnetic storms \citep{Gosling1990,Srivastava2004,RichardsonCane2012,Kilpua2012} causing communication and navigation system problems, damaging satellites and can even cause power outages \citep{Cannon2013}. The need for accurate predictions of CMEs, both CME arrival time and speed, is becoming increasingly important \citep{Owens2020}, because humankind, more than ever, depends on advanced technology.

Shortly after their eruption, CMEs can be observed in coronagraph images. Two of the few space-borne coronagraphs in operation are the Large Angle and Spectrometric Coronagraph (LASCO) C2 and C3 on-board the Solar and Heliospheric Observatory \citep[SoHO;][]{Brueckner1995}. SoHO is situated in a Lissajous orbit around Lagrange point 1 (L1), about 1.5 million km upstream of Earth in the Sun-Earth line. 

The launch of the Solar Terrestrial Relations Observatory \citep[STEREO;][]{Kaiser2008} twin-spacecraft mission in 2006 provided an unprecedented opportunity to observe CMEs from off the Sun-Earth line. The two spacecraft orbit the Sun slightly closer (STEREO Ahead; STA) and slightly further (STEREO Behind; STB) than Earth, leading to a separation of each spacecraft by about 22\textdegree\ per year from Earth in opposite directions. Both spacecraft are equipped with the In-situ Measurements of Particles and CME Transients \citep[IMPACT;][]{Luhmann2008} instrument package to measure solar wind speed, density and magnetic field and additionally host a suite of imagers, such as the COR1 and COR2 \citep{Howard2008} coronagraphs and the heliospheric imagers, HI1 and HI2 \citep{Eyles2009}. The wide-angle HI cameras provide observations of the heliosphere that allow us to track a CME from close to the Sun out to the orbit of Earth, particularly in the ecliptic plane. 

CMEs are optically thin structures that expand rapidly, and decreasing density lowers the line-of-sight integrated intensity in white-light data. As a consequence, the tracking of CME fronts and the interpretation of HI image data is difficult. 
Furthermore, the plane-of-sky assumption is not valid, and we must assume a certain longitudinal extent of the CME frontal shape. 

CMEs may be influenced by different phenomena in the heliosphere, e.g.~magnetic forces close to the Sun, high-speed solar wind streams, or by other CMEs \citep{Lugaz2012,KayOpher2015,Moestl2015}. The ambient solar wind can also affect the kinematic and morphological characteristics of CMEs \citep[e.g.][]{Gosling1990,Gopalswamy2000,Manoharan2004}. A CME originating at a speed much faster than the ambient solar wind speed is likely to experience deceleration while slow CMEs may accelerate during their propagation \citep{RichardsonCane2010,Manoharan2011}. Hence, not only the propagation direction but also the kinematics and shape of CMEs can be altered \citep[e.g.][]{Savani2010,Zuccarello2012,Liu2014,Rollett2014,Ruffenach2015,KayNievesChinchilla2020}. By tracking CMEs far out in the heliosphere, we get an understanding of their interaction with the ambient solar wind and co-rotating interaction regions.

Over the last decades, a vast number of CME prediction models have been developed. They include empirical models, e.g.~Effective Acceleration Model \citep[EAM;][]{Paouris2017}, which use relationships between observable parameters and the transit time. There are also drag-based models, (e.g.~DBM; \citealt{Vrsnak2013}, DBEM; \citealt{Dumbovic2018}, ANTEATR; \citealt{Kay2020}), that make use of physics-based equations and account for drag between the ambient solar wind and the CME. Other models make use of HI images, which require techniques to convert the measured elongation into radial distance. For example, the fixed phi fitting \citep[FPF;][]{Sheeley1999,Rouillard2008} technique considers a CME as a single point, propagating at a constant speed, and provides an estimate of the constant direction of the CME propagation relative to the observer from the apparent acceleration within a sequence of HI images. The harmonic mean fitting \citep[HMF;][]{Lugaz2010,Moestl2011} method is similar except that it describes a CME as a circle that remains attached to the Sun-center. The self-similar-expansion fitting \citep[SSEF;][]{LugazEtAl2010,Davies2012,MoestlDavies2013} technique describes a CME as a circle having an increasing radius as it propagates away from the Sun in such a way that it maintains a constant angular width. FPF and HMF are extremes of the SSEF technique with a half width of 0\textdegree\ and 90\textdegree, respectively. More sophisticated models combine both the drag-based approach and HI observations (e.g.~DBM fitting; \citealt{Zic2015}, Ellipse Evolution model based on HI observations, ELEvoHI; \citealt{Rollett2016, Amerstorfer2018}). Finally, numerical models, which are computational heavy, solve magnetohydrodynamic (MHD) equations (e.g., ENLIL; \citealt{Odstrcil2004}, EUHFORIA; \citealt{Pomoell2018}) simulating the ambient solar wind in the full heliosphere based on synoptic photospheric magnetic-field maps. CMEs are then injected into these models to provide predictions regarding the arrival time and arrival speed at different locations in the heliosphere.

As noted above, ELEvoHI aims to predict the arrival time and arrival speed of CMEs. The model assumes an elliptical shape for the CME front and incorporates the drag exerted by the ambient solar wind. Also, different sources of ambient solar wind speed (e.g. provided by numerical models) can serve as input to ELEvoHI \citep{Amerstorfer2021}. In its latest version, the model can be used with STEREO-A HI beacon mode data to provide near real-time CME arrival predictions. 

This study assesses ELEvoHI to evaluate arrival time and speed predictions of past CMEs using STEREO HI science-quality data. We perform ELEvoHI ensemble predictions for 12 CMEs, where each CME is modeled using input data from STA and STB, separately. In an idealized case, in which a CME with an elliptical front propagates in an ambient solar wind that is constant in space and time, one would expect to get similar results for the arrival time and arrival speed from the two different vantage points. Instead of inferring the propagation directions of the events under study from HI images (e.g.~FPF, SSEF), as was done by \cite{Amerstorfer2018}, we make use of coronagraph images and perform Graduated Cylindrical Shell \citep[GCS;][]{Thernisien2006, Thernisien2009} reconstruction for each CME based on multi-vantage point coronagraph data. Additionally, we apply a combination of the Wang-Sheeley-Arge~\citep{arge03} and the Heliospheric Upwind eXtrapolation~\citep{riley11b,Owens2017} model \citep[WSA/HUX model combination;][]{reiss19,Reiss2020ApJ} to get an estimate of the ambient solar wind conditions in the heliosphere through which the CME propagates. With the additional information about the propagation direction of the CME and the modeled ambient solar wind, ELEvoHI is more likely to give better arrival time and arrival speed predictions.

In Section~\ref{Sec:Data}, we describe our data selection process, including the data products, and list all of the studied CMEs. Section~\ref{sec:Methods} deals with the ELEvoHI setup and how the input data required by the model is obtained. In Section~\ref{sec:Results}, we present our results and give reasons for the difference in the model predictions based on STA and STB input data. The discussion and further implementations of the model are included in Section~\ref{Sec:Summary}.

\section{Data Preparation}\label{Sec:Data}

\begin{table}
\begin{threeparttable}
\caption{List of selected CMEs. \textit{ID} and \textit{Date} correspond to the unique identifier and the time of the first appearance of the CME in HI1 imagery, from the \href{https://www.helcats-fp7.eu}{HELCATS} catalog, for STA and STB spacecraft. \textit{ICMECAT ID} is the identifier of the interplanetary coronal mass ejection (ICME) from an updated version of the \href{https://helioforecast.space/icmecat}{HELCATS ICMECAT} \citep{Moestl2017}, \textit{ICME date} is the start time of the detected ICME and $v_{\mathrm{ICME}}$ is the measured in situ arrival speed obtained from the HELCATS ICMECAT. }
\label{tab:selCMEs} 
\scriptsize
\begin{tabular}{c|c|c|c|c|c|c|c}
\hline
Nr. & ID STA & Date STA & ID STB & Date STB & ICMECAT ID & ICME date & $v_{\mathrm{ICME}}$ \\ 
 & & & & & & & [km~s$^{-1}$] \\ 
\hline
1 & \makecell{HCME\_A\_\_\\20100203\_01} & \makecell{2010-02-03\\14:49} & \makecell{HCME\_B\_\_\\20100203\_01} & \makecell{2010-02-03\\20:49} & \makecell{ICME\_Wind\_\\NASA\_20100207\_01} & \makecell{2010-02-07\\18:04\tnote{b)}} & 406$\pm$2\\ \hdashline
2 & \makecell{HCME\_A\_\_\\20100319\_01} & \makecell{2010-03-19\\22:09} & \makecell{HCME\_B\_\_\\20100319\_01} & \makecell{2010-03-19\\20:09} & \makecell{ICME\_Wind\_\\MOESTL\_20100323\_01} & \makecell{2010-03-23\\22:29\tnote{c)}} & 292$\pm$12 \\ \hdashline
3 & \makecell{HCME\_A\_\_\\20100403\_01} & \makecell{2010-04-03\\12:09} & \makecell{HCME\_B\_\_\\20100403\_01} & \makecell{2010-04-03\\12:09} & \makecell{ICME\_Wind\_\\NASA\_20100405\_01} & \makecell{2010-04-05\\07:55\tnote{a)}} & 734$\pm$18 \\ \hdashline
4 & \makecell{HCME\_A\_\_\\20100408\_01} & \makecell{2010-04-08\\06:49} & \makecell{HCME\_B\_\_\\20100408\_01} & \makecell{2010-04-08\\07:29} & \makecell{ICME\_Wind\_\\NASA\_20100411\_01} & \makecell{2010-04-11\\12:20\tnote{a)}} & 432$\pm$17 \\ \hdashline
5 & \makecell{HCME\_A\_\_\\20100523\_01} & \makecell{2010-05-23\\22:09} & \makecell{HCME\_B\_\_\\20100524\_01} & \makecell{2010-05-24\\00:09} &  \makecell{ICME\_Wind\_\\NASA\_20100528\_01} & \makecell{2010-05-28\\01:52\tnote{a)}} & 370$\pm$10 \\ \hdashline
6 & \makecell{HCME\_A\_\_\\20101026\_01} & \makecell{2010-10-26\\15:29} & \makecell{HCME\_B\_\_\\20101026\_01} & \makecell{2010-10-26\\16:10} & \makecell{ICME\_Wind\_\\MOESTL\_20101030\_01} & \makecell{2010-10-30\\09:15\tnote{b)}} & 380$\pm$9 \\ \hdashline
7 & \makecell{HCME\_A\_\_\\20110130\_01} & \makecell{2011-01-30\\20:09} & \makecell{HCME\_B\_\_\\20110130\_01} & \makecell{2011-01-30\\18:49} & \makecell{ICME\_Wind\_\\MOESTL\_20110204\_01} & \makecell{2011-02-04\\01:50\tnote{a)}} & 375$\pm$9 \\ \hdashline
8 & \makecell{HCME\_A\_\_\\20110214\_02} & \makecell{2011-02-14\\22:49} & \makecell{HCME\_B\_\_\\20110214\_02} & \makecell{2011-02-14\\22:09} & \makecell{ICME\_Wind\_\\MOESTL\_20110218\_01} & \makecell{2011-02-18\\00:48\tnote{a)}} & 493$\pm$25 \\ \hdashline
9 & \makecell{HCME\_A\_\_\\20110906\_02} & \makecell{2011-09-06\\23:29} & \makecell{HCME\_B\_\_\\20110907\_01} & \makecell{2011-09-07\\03:29} & \makecell{ICME\_Wind\_\\MOESTL\_20110909\_01} & \makecell{2011-09-09\\11:46\tnote{a)}} & 417$\pm$20 \\ \hdashline
10 & \makecell{HCME\_A\_\_\\20120123\_01} & \makecell{2012-01-23\\04:49} & \makecell{HCME\_B\_\_\\20120123\_01} & \makecell{2012-01-23\\05:29} & \makecell{ICME\_Wind\_\\MOESTL\_20120124\_01} & \makecell{2012-01-24\\14:36\tnote{a)}} & 613$\pm$36 \\ \hdashline
11 & \makecell{HCME\_A\_\_\\20120614\_01} & \makecell{2012-06-14\\16:09} & \makecell{HCME\_B\_\_\\20120614\_01}  & \makecell{2012-06-14\\16:09} & \makecell{ICME\_Wind\_\\MOESTL\_20120616\_01} & \makecell{2012-06-16\\19:34\tnote{a)}} & 489$\pm$29 \\ \hdashline
12 & \makecell{HCME\_A\_\_\\20120712\_02} & \makecell{2012-07-12\\18:49} & \makecell{HCME\_B\_\_\\20120712\_01}  & \makecell{2012-07-12\\18:09} & \makecell{ICME\_Wind\_\\MOESTL\_20120714\_01} & \makecell{2012-07-14\\17:38\tnote{a)}} & 615$\pm$37 \\
\hline
\end{tabular}

\begin{tablenotes}
\item[a)] shock arrival time
\item[b)] time of density enhancement
\item[c)] time of the magnetic flux rope
\end{tablenotes}
\end{threeparttable}
\end{table}    
We select a period between February 2010 and July 2012 during which the STEREO spacecraft had a separation angle from Earth of about 65\textdegree\ to 120\textdegree\ respectively, from which we study 12 CMEs.
The \href{https://www.helcats-fp7.eu}{HELCATS} HICAT CME catalog lists about 700 entries over this time range \citep{Harrison2018}.
However, our list is constrained to 12 events, since the CMEs have to:
\begin{itemize}
    \item[1)] be observed by HI on both STA and STB spacecraft \citep[as listed in the HIJoinCAT;][]{Barnes2020}
    \item[2)] propagate close to the ecliptic plane,
    \item[3)] have a corresponding in situ signature at Earth,
    \item[4)] be able to be tracked unambiguously in time-elongation maps.
\end{itemize}

Table \ref{tab:selCMEs} contains the list of selected CMEs with their unique identifier and the time of their first observation in HI1 images (according to the \href{https://www.helcats-fp7.eu}{HELCATS} catalog Version 6). The interplanetary CME (ICME) times and speeds are taken from version 2.0 of the HELCATS ICMECAT catalog \citep[][see also the links in the data section]{Moestl2020}. The ICMECAT assimilates ICME catalogs from different spacecraft into one consistent list, and was first published in \cite{Moestl2017}.
The ICME date as observed by the Wind spacecraft is defined by the shock arrival time, or, if no shock is present, the start of a density enhancement in front of the magnetic flux rope (MFR). If neither is observed, the ICME start time is taken as the start time of the MFR. The corresponding ICME speed is the mean proton bulk speed of either the sheath region, the density enhancement ahead of the MFR, or the speed of the MFR itself. The spread in the speed over the given interval for each event is indicated in Table \ref{tab:selCMEs} by a standard deviation. For Table \ref{tab:selCMEs}, some times in the ICMECAT were originally taken from the Wind ICME catalog \citep{Nieves_Chinchilla2018}, while other events that were not present in the Wind catalog were added by \cite{Moestl2020} to the HELCATS ICMECAT.

To run ELEvoHI, we make use of several data products. Most important are images from HI onboard STEREO. The HI instrument on each STEREO spacecraft consists of two white-light wide-angle imagers, HI1 and HI2. HI1 has a field-of-view (FOV) extending from 4\textdegree\ -- 24\textdegree\ elongation (angle from Sun center) in the ecliptic and HI2 has an angular FOV extending from 18.8\textdegree\ -- 88.8\textdegree\ elongation in the ecliptic. 
The nominal cadence of the HI1 and HI2 science data is 40 minutes and 120 minutes, respectively. The science image bin size is 70 arc sec for HI1 and 4 arc min for HI2. For the additional input parameters to ELEvoHI, we developed the Ecliptic cut Angles from GCS for ELEvoHI tool (EAGEL, see Section~\ref{sec:EAGEL}). EAGEL ideally uses coronagraph images from STEREO COR1/COR2 and from LASCO C2/C3 onboard SoHO, but  images from at least two different viewpoints are required. The FOV of COR1 ranges from 1.4 -- 4
~R$_{\odot}$ and COR2, from 2 -- 15~R$_{\odot}$, while C2 has a FOV of 1.5 -- 6~R$_{\odot}$ and C3, 3.7 -- 30~R$_{\odot}$ (all quoted in the plane-of-sky). The cadence of the coronagraph science images is about 15 minutes. 


\section{Methods} \label{sec:Methods}
\subsection{EAGEL (Ecliptic cut Angles from GCS for ELEvoHI)} \label{sec:EAGEL}

\begin{figure}[htp]
\centering
\includegraphics[width=1\linewidth]{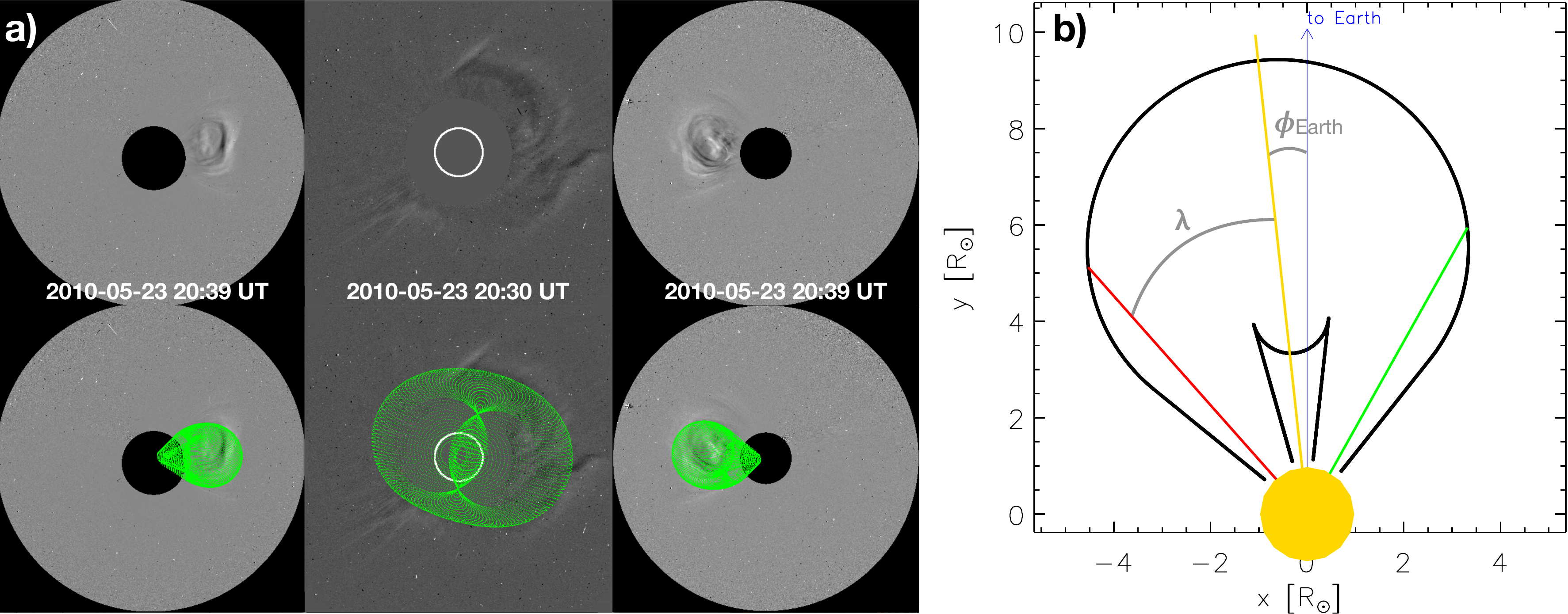}
\caption{GCS reconstruction (left) and ecliptic cut of the wireframe (right) for event $\#$5. a) Top row from left to right: STB/COR2, LASCO/C2, STA/COR2. Bottom row: same as top row but with the GCS wire frame overlaid. b) Ecliptic cut (black) of the GCS wire frame. Red and green lines show the boundaries selected by either EAGEL or the user. The yellow line defines the ecliptic propagation direction with respect to Earth, $\phi_{Earth}$, of the CME. The half width, $\lambda$, is the angle between one boundary and $\phi_{Earth}$. The blue arrow indicates the direction to Earth.} \label{fig:GCS_wire}
\end{figure}

In this section, we present a newly developed Interactive Data Language (IDL\textsuperscript{TM}) tool called EAGEL (\textbf{E}cliptic cut \textbf{A}ngles from \textbf{G}CS for \textbf{EL}EvoHI). EAGEL allows the user to determine the propagation direction, $\phi$, and the half width, $\lambda$, 
within the ecliptic plane, based on GCS reconstruction of a CME. 
To perform GCS reconstruction, coronagraph images from at least two vantage points (STEREO and/or LASCO) are required. EAGEL provides the routines to download the required coronagraph images, combines all the functions to perform GCS reconstruction, and produces a cut in the ecliptic plane. Standard pre-processing of the images is implemented in EAGEL to make the CME features clearly visible to the user, who can decide between using background-subtracted, running-difference, and base-difference images. The user can then perform GCS reconstruction using the IDL SolarSoft procedure \texttt{rtsccguicloud}. The top row of Figure~\ref{fig:GCS_wire}a shows the coronagraph images (from left to right: STB/COR2, LASCO/C2, STA/COR2) for event $\#$5. The bottom row additionally shows the GCS wire frame (green mesh). In its current version, ELEvoHI is a 2D prediction model giving results only in the ecliptic plane. Therefore, EAGEL calculates the ecliptic part of the GCS wire frame and selects the boundaries of the ecliptic cut (see red and green line in Figure~\ref{fig:GCS_wire}b). The boundaries are defined to be the outermost points of each side of the ecliptic cut with respect to the apex direction from GCS reconstruction. This gives $\lambda$ and $\phi$, where the latter is defined to be exactly in between the two boundaries. A plot is shown to the user (Figure~\ref{fig:GCS_wire}b) and, if needed, the boundaries can be changed manually. Once the user approves the selection, $\lambda$ and $\phi$ relative to Earth and to the two STEREO spacecraft are stored and can be used by ELEvoHI.

In Table \ref{tab:GCS}, we list the time (Date) of the STEREO coronagraph images used to get $\lambda$ and $\phi$ for each event. EAGEL then selects the SoHO coronagraph images closest in time to the quoted date. Each CME is fitted once based on the three different viewpoints (STA, STB, LASCO). However, for event $\#$1 no LASCO data is available, so GCS reconstruction is based on STEREO images only. For all the events, the times of the images are selected in such a way that the CME front is clearly visible in the coronagraph images of all the viewpoints. Furthermore, we try to fit the CME at times when the front is already far out in both STA and STB COR2 images. Table \ref{tab:GCS} additionally contains the GCS parameters (\textit{Lon, Lat, TA, AR, HA}). Also the half width, $\lambda$, and the CME ecliptic propagation angle, $\phi$, relative to Earth ($\phi_{\mathrm{Earth}}$) and relative to the two STEREO spacecraft ($\phi_{\mathrm{STA}}$ and $\phi_{\mathrm{STB}}$) obtained from EAGEL are given. \textit{Lon} is the longitude (here given in Stonyhurst coordinates) and \textit{Lat} the latitude of the apex of the idealized hollow croissant shaped model. The tilt angle (\textit{TA}) defines the tilt of the croissant, calculated with respect to the solar equator. The half angle (\textit{HA}) represents the angle between the center of the footpoints and the aspect ratio (\textit{AR}) is the ratio of the CME size in two orthogonal directions.

When comparing $Lon$ (longitude from GCS reconstruction) and $\phi_{\mathrm{Earth}}$ (longitude relative to Earth from the ecliptic cut), it can be seen that the propagation direction obtained from the ecliptic cut is quite comparable to (within 5\textdegree\ of) the propagation direction from the GCS reconstruction. Only for events $\#$6 and $\#$10 we find differences of about 30\textdegree\ and 10\textdegree, respectively. The reason can be found in the combination of low/high latitude and large tilt angle. Therefore, the part within the ecliptic plane does not correspond well to the main propagation direction resulting from GCS reconstruction  for these two CMEs.

\begin{table}[htbp]
\caption{GCS parameter obtained from fitting the hollow croissant shape to the STEREO and SoHO coronagraph images and EAGEL results. \textit{Date}: time set in EAGEL to perform the reconstruction. \textit{Lon}: longitude (Stonyhurst coordinates), \textit{Lat}: latitude, \textit{TA}: tilt angle, \textit{AR}: aspect ratio, \textit{HA}: half angle from GCS. The remaining values are based on the ecliptic cut from EAGEL: $\lambda$: half width of the CME, $\phi_{\mathrm{Earth}}$, $\phi_{\mathrm{STA}}$, $\phi_{\mathrm{STB}}$: propagation direction with respect to Earth, STA, STB, respectively.} \label{tab:GCS}
\centering
\scriptsize
\begin{tabular}{cc!{\vrule width 0.08em}ccccc!{\vrule width 0.08em}cccc}
\hline
& & \multicolumn{5}{c!{\vrule width 0.08em}}{GCS parameter} & \multicolumn{4}{|c|}{EAGEL results} \\
\hline
Nr. & Date & Lon & Lat & TA & AR & HA & $\lambda$ & $\phi_{\mathrm{Earth}}$ & $\phi_{\mathrm{STA}}$ & $\phi_{\mathrm{STB}}$\\
& & [\textdegree] & [\textdegree] & [\textdegree] & & [\textdegree] & [\textdegree] & [\textdegree] & [\textdegree] & [\textdegree] \\
\hline 
1 & 2010-02-03 15:54 & 355 & -17 & -1 & 0.33 & 30 & 36 & -4 & 67 & 68 \\
2 & 2010-03-19 17:39 & 23 & -12 & -7 & 0.29 & 19 & 30 & 22 & 44 & 93 \\
3 & 2010-04-03 12:39 & 7 & -19 & 15 & 0.39 & 30 & 38 & 9 & 58 & 81 \\
4 & 2010-04-08 06:39 & 1 & -10 & -20 & 0.28 & 30 & 31 & -2 & 70 & 69 \\
5 & 2010-05-23 20:39 & 6 & 2 & -15 & 0.48 & 18 & 35 & -6 & 65 & 76 \\
6 & 2010-10-26 14:39 & 18 & -35 & -28 & 0.51 & 30 & 18 & -11 & 95 & 69 \\
7 & 2011-01-30 21:24 & 351 & -18 & -20 & 0.33 & 12 & 24 & -11 & 97 & 82 \\
8 & 2011-02-15 04:08 & 10 & -10 & 27 & 0.87 & 29 & 49 & 10 & 77 & 104 \\
9 & 2011-09-06 23:39 & 29 & 20 & -90 & 0.49 & 30 & 26 & 29 & 74 & 124 \\
10 & 2012-01-23 04:39 & 19 & 41 & 64 & 0.77 & 55 & 37 & 9 & 99 & 123 \\
11 & 2012-06-14 14:54 & 360 & -28 & 11 & 0.90 & 30 & 53 & 1 & 116 & 117 \\
12 & 2012-07-12 17:54 & 8 & -12 & 68 & 0.46 & 30 & 26 & 14 & 106 & 129 \\
\hline
\end{tabular}
\end{table}

\subsection{WSA/HUX model}\label{HuxModel}

In the following paragraph, we summarize the main characteristics of the numerical framework used here for modelling the physical conditions in the evolving ambient solar wind flow. For this study, we make use of the framework shown in \cite{reiss19,Reiss2020ApJ}, but the components of this framework were developed by \cite{wang95}, \cite{arge03}, \cite{riley11b}, and \cite{Owens2017}. Specifically, we use magnetic maps of the photospheric field from Global Oscillation Network Group (GONG) provided by the National Solar Observatory (NSO) as input to magnetic models of the solar corona. Using the Potential Field Source Surface model~\citep[PFSS;][]{altschuler69, schatten69} and the Schatten current sheet model~\citep[SCS;][]{schatten71} we compute the global coronal magnetic field topology. While the PFSS model attempts to find the potential magnetic field solution in the corona with an outer boundary condition that the field is radial at the source surface at 2.5~R$_{\odot}$, the SCS model in the region between 2.5 and 5~R$_{\odot}$ accounts for the latitudinal invariance of the radial magnetic field as observed by Ulysses \citep[][]{wang95}. From the global magnetic field topology, we calculate the solar wind conditions near the Sun using the established Wang-Sheeley-Arge (WSA) model.
To map the solar wind solutions from near the Sun to Earth, we use the Heliospheric Upwind eXtrapolation model (HUX). This model simplifies the fluid momentum equation as much as possible, by neglecting the pressure gradient and the gravitation term in the fluid momentum equations as proposed by \cite{riley11b}. The HUX model solutions match the dynamical evolution explored by global heliospheric MHD codes fairly well while having low processor requirements. An example of the ambient solar wind, modeled by WSA/HUX combination, is shown in Figure~\ref{fig:WSA_BGSW}.

\subsection{ELEvoHI ensemble modeling}\label{sec:ELEvoHI}
\href{https://zenodo.org/record/3873420}{ELEvoHI} uses HI time-elongation profiles of CME fronts and assumes an elliptical shape for those fronts to derive their interplanetary kinematics \citep[detailed information about the underlying Ellipse Conversion method can be found in ][]{Rollett2016}. The tracking of each CME was done manually using ecliptic time-elongation maps \citep[j-maps;][]{Sheeley1999, Davies2009}, generated by extracting ecliptic data from STA and STB HI images. 
Transients, like CMEs, appear as a bright feature in the j-maps. To extract the time-elongation profiles, we use the \href{https://hesperia.gsfc.nasa.gov/ssw/stereo/secchi/idl/jpl/satplot/SATPLOT_User_Guide.pdf}{SATPLOT} tool implemented in IDL\textsuperscript{TM} SolarSoft. It allows any user to measure the elongation, which is defined as the angle between the Sun - observer (STA or STB) line and the CME front. ELEvoHI converts the resulting time-elongation profiles to time-distance profiles, assuming an elliptic frontal shape using the ELEvoHI built-in procedure ELlipse Conversion \citep[ELCon;][]{Rollett2016}. 

\begin{figure}[htbp]
\centering
\includegraphics[width=0.7\linewidth]{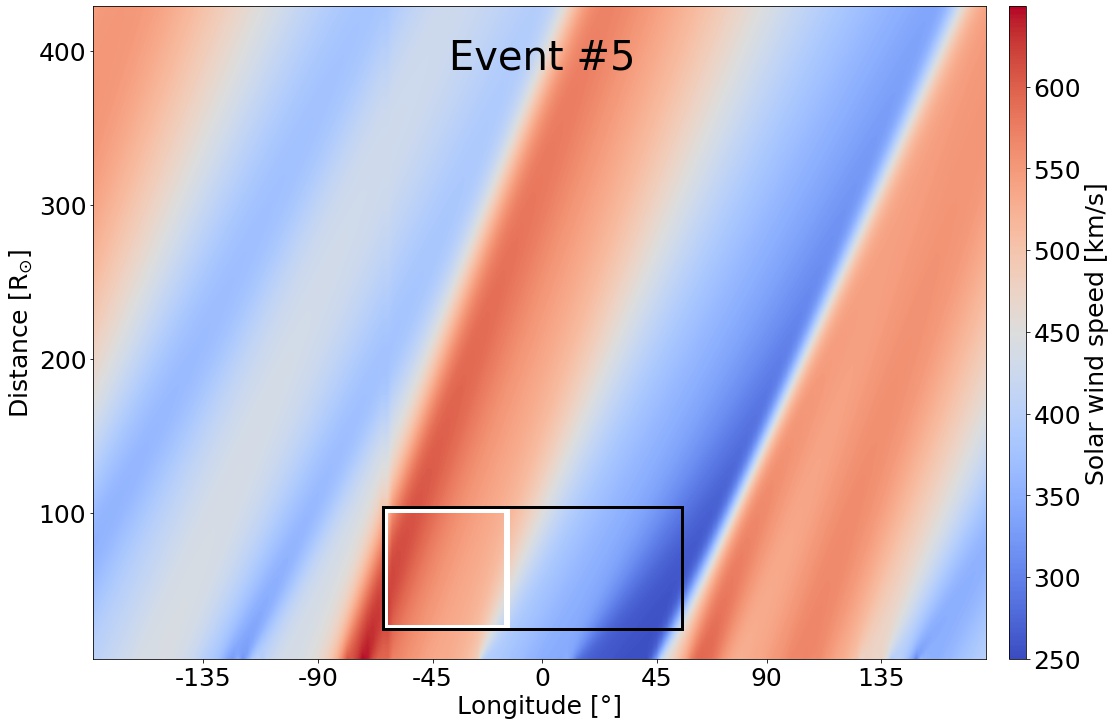}
\caption{Ambient solar wind speed provided by the WSA/HUX model for event $\#$5. The white box defines the area that is used to calculate an estimate of the ambient solar wind speed for the ensemble member of ELEvoHI corresponding to the minimum propagation direction ($\phi_{\mathrm{STA}}=56$\textdegree) with the maximum half width ($\lambda=50$\textdegree). The black box indicates the total area based on all the ensemble members of ELEvoHI for this event. The longitude of 0\textdegree\ corresponds to the longitude of Earth.} \label{fig:WSA_BGSW}
\end{figure}

ELEvoHI accounts for the effect of the drag force exerted by the ambient solar wind, which is incorporated in the model. The drag force is an essential factor influencing the dynamic evolution of CMEs in the heliosphere. Within ELEvoHI, the time-distance track is fitted using a drag-based equation based on the drag-based model (DBM) given in \cite{Vrsnak2013}.
The user has to define the start- and end point for the DBM fit (usually around $30-100$~R$_{\odot}$) in the time-distance profile. In order to account for the de-/acceleration of the CME due to drag, an estimate of the ambient solar wind speed is needed. Here we make use of the WSA/HUX model (see Section~\ref{HuxModel}). It provides the ambient solar wind conditions for a full Carrington rotation (see Figure~\ref{fig:WSA_BGSW}). We only consider the part of the full map according to the start- and end-point selected by the user, and the CME propagation direction and half width from EAGEL. 
From this area surrounded by the white box in Figure~\ref{fig:WSA_BGSW}, we take the median of the solar wind speed and define the uncertainties to be $\pm$100~km~s$^{-1}$, based on a study by \cite{Reiss2020ApJ}. They considered nine years (mid 2006 to mid 2015) and report a mean absolute error of the WSA solar wind speed prediction with respect to the in situ speed of 91~km~s$^{-1}$.
The obtained ambient solar wind speed with its uncertainty is split into steps of 25~km~s$^{-1}$, which gives nine different input speeds to ELEvoHI. For each of the nine input speeds DBM fitting is performed. ELEvoHI then selects the combination of drag parameter and ambient solar wind speed that best fits the time-distance profile for each ensemble member \citep[for a detailed description see][]{Rollett2016}. The selected drag parameter and solar wind speed is assumed to be valid for the full CME front during the propagation in the heliosphere.

Since ELEvoHI is a 2D model, we are only interested in the propagation of a CME in the ecliptic plane. $\phi$ and $\lambda$, in this plane, are provided by EAGEL (see Section~\ref{sec:EAGEL}). The inverse ellipse aspect ratio, $f$, defines the shape of the assumed CME front in the ecliptic plane, where $f=1$ represents a circular front, while $f<1$ corresponds to an elliptical CME front (with the semi-major axis perpendicular to the propagation direction)

To run ELEvoHI in ensemble mode, we vary $\phi$, $\lambda$, and $f$. A details description can be found in \cite{Amerstorfer2018} and the code is available \href{https://zenodo.org/record/3873420}{online} (see Section~\ref{sec:DataSources}). $\phi$ and $\lambda$ vary over a range of $\pm10$\textdegree\ from their values obtained from EAGEL, with a step size of $2$\textdegree\ and $5$\textdegree, respectively.
This range is defined based on a study by \cite{Mierla2010}, who report an uncertainty in the parameters when different users manually perform GCS reconstruction. Note that the propagation direction and the half width obtained from EAGEL are rounded to even numbers and to whole tens, respectively. For $f$ we set a fixed range from $0.7-1.0$ ($0.1$ step size). Thus we get a total of 220 ensemble members for one ELEvoHI event (i.e.~11 values of $\phi$, 5 values of $\lambda$ and 4 values of $f$). For each ensemble member we select a different sector from the ambient solar wind provided by the WSA/HUX model combination according to the propagation direction, half width, start- and end-point. In Figure~\ref{fig:WSA_BGSW}, the WSA/HUX model results for event $\#5$ are shown. The white box indicates the area from which the ambient solar wind speed for one individual run of ELEvoHI is computed. Shown is the area for the minimum propagation direction, $\phi_{\mathrm{STA}}$ of 56\textdegree\ with a $\lambda$ of 50\textdegree. For each ensemble member the area surrounded by the white box is slightly different according to $\phi$ and $\lambda$. The black box plotted indicates the total area based on all ELEvoHI ensemble members for this event. 

Running ELEvoHI in ensemble mode enables us to calculate a mean and a median predicted CME arrival time and also to define an uncertainty. In addition, we can give a probability for whether a CME is likely to hit Earth or not. When all of the 220 ensemble members predict an arrival at Earth, we assume the predicted likelihood of an Earth hit to be 100\%.

\section{Results} \label{sec:Results}

\begin{table}
\caption{List of predicted median arrival times (\textit{Date}) and the standard deviation (\textit{SD}) based on STA and STB observations, respectively. \textit{STA - STB} gives the difference between the predicted median arrival times. $v$ is the predicted median arrival speed with the standard deviation and $v_{\mathrm{STA-STB}}$ is the difference in arrival speed between STA and STB predictions}\label{tab:ArrTimes}
\centering
\scriptsize
\begin{tabular}{c|c|c|c|c|c|c|c|c}
\hline
Nr. & Date STA & SD$_{\mathrm{STA}}$ & Date STB & SD$_{\mathrm{STB}}$ & STA - STB & $v_{\mathrm{STA}}$ & $v_{\mathrm{STB}}$ & $v_{\mathrm{STA-STB}}$ \\
& & [h] & & [h] & [h] & [km~s$^{-1}$] & [km~s$^{-1}$] & [km~s$^{-1}$] \\
\hline 
1 & \makecell{2010-02-07\\11:24} & 1.5 & \makecell{2010-02-07\\20:24} & 2.1 & -9.0 & $455\pm17$ & $395\pm11$ & 60 \\ \hdashline
2 & \makecell{2010-03-24\\07:17} & 9.1 & \makecell{2010-03-24\\16:40} & 4.1 & -9.5 & $401\pm32$ & $351\pm11$ & 50 \\ \hdashline
3 & \makecell{2010-04-05\\13:23} & 2.5 & \makecell{2010-04-05\\16:06} & 0.4 & -2.7 & $649\pm37$ & $625\pm5$ & 24 \\ \hdashline
4 & \makecell{2010-04-11\\16:07} & 0.6 & \makecell{2010-04-12\\00:12} & 5.1 & -8.1 & $443\pm6$ & $391\pm33$ & 52 \\ \hdashline
5 & \makecell{2010-05-27\\17:36} & 1.9 & \makecell{2010-05-28\\02:26} & 1.2 & -8.8 & $455\pm9$ & $407\pm9$ & 48 \\ \hdashline
6 & \makecell{2010-10-30\\11:24} & 1.4 & \makecell{2010-10-30\\04:43} & 7.1 & 6.7 & $432\pm7$ & $476\pm45$ & -44 \\ \hdashline
7 & \makecell{2011-02-04\\01:08} & 2.4 & \makecell{2011-02-03\\22:24} & 7.3 & 4.5 & $387\pm9$ & $446\pm34$ & -59 \\ \hdashline
8 & \makecell{2011-02-18\\06:22} & 2.8 & \makecell{2011-02-18\\10:34} & 6.1 & -4.3 & $478\pm18$ & $407\pm50$ & 71 \\ \hdashline
9 & \makecell{2011-09-10\\18:55} & 14.9 & \makecell{2011-09-10\\09:48} & 5.4 & 9.1 & $396\pm46$ & $430\pm18$ & -34 \\ \hdashline
10 & \makecell{2012-01-24\\17:49} & 4.0 & \makecell{2012-01-24\\13:29} & 3.6 & 4.3 & $793\pm103$ & $982\pm150$ & -189 \\ \hdashline
11 & \makecell{2012-06-16\\15:47} & 3.8 & \makecell{2012-06-16\\07:53} & 5.2 & 7.9 & $712\pm72$ & $749\pm143$ & -37 \\ \hdashline
12 & \makecell{2012-07-14\\22:16} & 4.9 & \makecell{2012-07-14\\18:53} & 3.7 & 3.5 & $658\pm80$ & $579\pm28$ & 89 \\
\hline
\end{tabular}
\end{table}

We perform ELEvoHI ensemble modeling for 12 CMEs between February 2010 to July 2012 (see Table \ref{tab:selCMEs}) and compare the predicted arrival times based on STA and STB HI observations with each other. The CMEs propagated close to the ecliptic plane and showed clear in situ signatures at L1. A prerequisite for the chosen CMEs was that the CMEs could be tracked unambiguously in both STA and STB HI j-maps.

In Table \ref{tab:ArrTimes}, we list the predicted ensemble median arrival times and speeds with their standard deviation for each CME under study. It further contains the difference between the predictions from the two vantage points. 
We find that the predicted arrival times for STA and STB can deviate by up to 9.5~hrs while the mean difference is 6.5~hrs. The mean difference in the arrival speed is 63~km~s$^{-1}$, with an exceptionally large discrepancy of 189~km~ s$^{-1}$ for event $\#10$.

The largest arrival time differences are found for events $\#2$ and $\#9$. The arrival probability, based on the number of ensemble members that are predicted to hit Earth, is 79\% for event $\#2$ and only 56\% for event $\#9$. According to their relatively large angle of propagation with respect to the Sun-Earth line, the CMEs $\#2$ ($\phi_{\mathrm{Earth}} = 22$\textdegree, $HA = 30$\textdegree) and $\#9$ ($\phi_{\mathrm{Earth}} = 30$\textdegree, $HA = 30$\textdegree) are considered as "flank hits". In such cases, ELEvoHI tends to predict the CME arrival time to be later than detected in-situ. The reason may be found in the assumed circular CME front for $f = 1.0$. For future versions of ELEvoHI, we will consider different approaches to tackle such extreme delays for flank encounters e.g. by changing the values for $f$ (from 0.4 to 0.7). \cite{Braga2020} found a value of $f \sim 0.6$ for most of the CMEs in their study.

Event $\#$11 occurred on June 14, 2012 and was studied e.g.~by \cite{Kubicka2016} who report two preceding CMEs. However, the WSA/HUX model does not provide the ambient solar wind conditions with preceding CMEs included and is therefore most probably not suitable for interaction events. The events $\#1$, $\#4$, and $\#5$ also show large differences in the predicted arrival times based on STA and STB observations. However, these differences are most certainly related to large variance in the modeled ambient solar wind speeds that are used as input to ELEvoHI (see Section~\ref{sec:ambientSW} and Figure~\ref{fig:AllSolarWinds} and \ref{fig:BGSW_ArrDiff}).

\begin{figure}[htbp]
\centering
\includegraphics[width=0.8\linewidth]{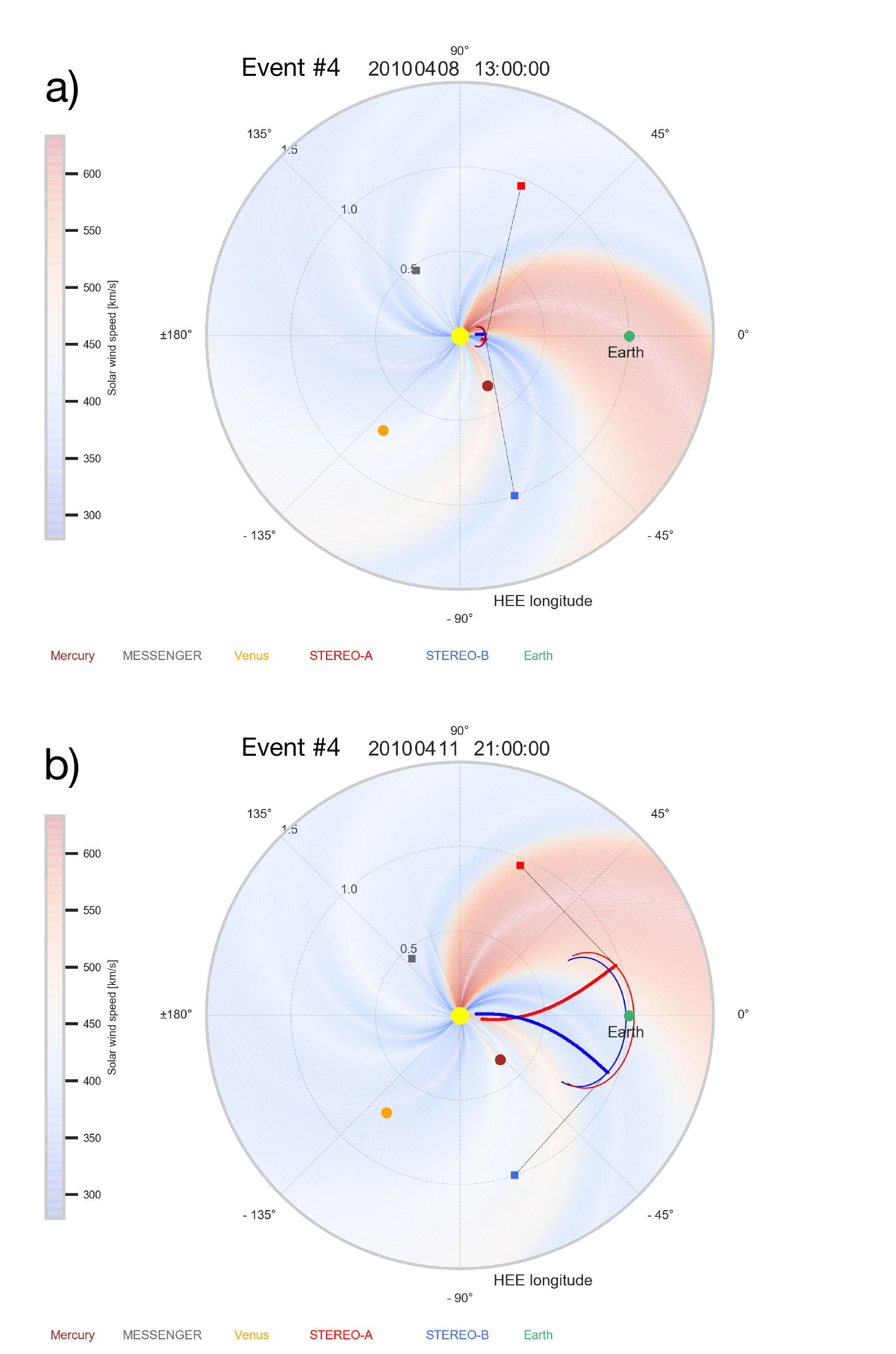}
\caption{Two snapshots of the CME propagation for one ensemble member based on STA (red) and STB (blue) observations for event $\#4$. The ambient solar wind is computed using the WSA/HUX model combination. The elliptical CME fronts from one ensemble member based on STA and STB observations are shown in red and blue, respectively. The thick curved lines in red and blue show the intercept of the idealized elliptical front of the CME and the tangent (gray lines) for each time step over the course of the simulation for STA and STB, respectively. Link to the \href{https://figshare.com/articles/media/20110130_AB_tangent_movie/13077608}{movie} (\href{https://figshare.com/articles/media/20110130_AB_tangent_movie/13077608}{https://figshare.com/articles/media/20110130\_AB\_tangent\_movie/13077608}).
} \label{fig:MovieSnapshot}
\end{figure}

\begin{figure}[htbp]
\centering
\includegraphics[width=0.8\linewidth]{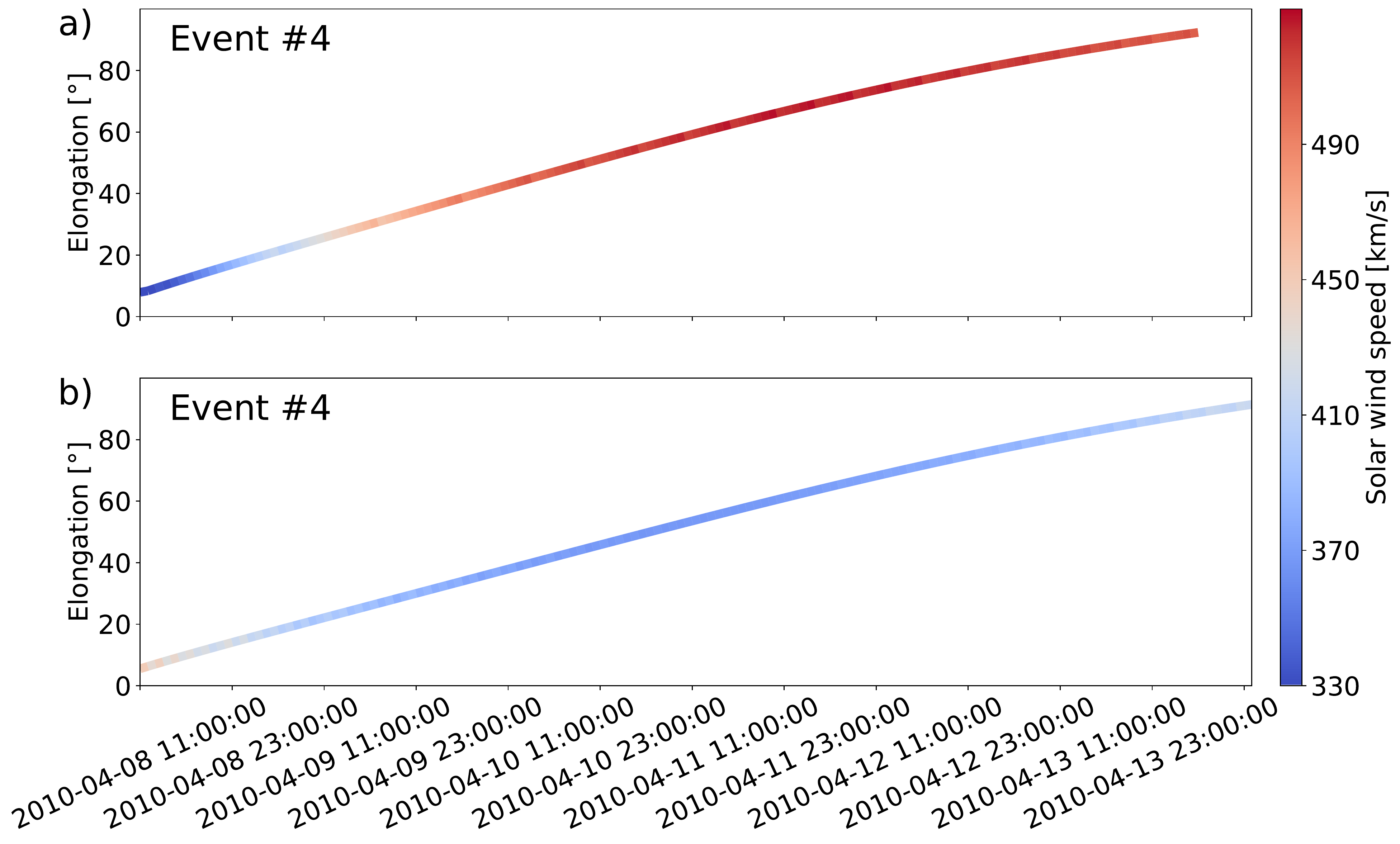}
\caption{Ambient solar wind speed at the tangent points for event $\#4$. Plotted are time series of the elongation angles of the tangent points as seen from STA (top panel) and STB (bottom panel) colour-coded according to the speed of the ambient solar wind at that tangent point.} \label{fig:STASTBElon}
\end{figure}

\subsection{Tracking different parts of the CME front}
It is important to keep in mind that different parts of the CME front are tracked in STA and STB HI images. This leads to different input conditions to the ELEvo propagation model for STA and STB. ELEvoHI is designed to take HI tracks for the same CME from different viewpoints. Ideally, predictions should give the same CME speed and direction in both cases. One problem is, however, that the CME is not behaving as a single coherent entity, but is instead moving with different speeds at different longitudes \citep{Owens2017Nat}, which is not incorporated within ELCon nor in any other HI conversion method (e.g. SSE, FP, and HM).

Figure~\ref{fig:MovieSnapshot} presents two snapshots of a \href{https://figshare.com/articles/20110130_AB_ensemble_movie_mp4/12179919}{movie} for event $\#4$, with the ambient solar wind provided by WSA/HUX model combination and the positions of various spacecraft and planets. The elliptical CME fronts from one ensemble member based on STA and STB observations are shown in red and blue, respectively. 
The gray lines from the two STEREO spacecraft to the elliptical CME fronts are plotted. These tangents correspond to the elongations of the leading edge of the CME at these times. At the end of these lines, we add a point, which is the 'tangent point' at each time step. Over the course of the simulation, these points trace out curved lines, in red and blue for STA and STB, respectively. 
From Figure~\ref{fig:MovieSnapshot}, it is obvious that, in the near-Sun part of the HI FOV, the observed leading edge is close to the apex of the idealized CME front for both STEREO spacecraft. As the CME propagates, the tangent point, i.e.~the part of the CME with the greatest elongation seen by STA and STB progressively moves out to the flanks of the ellipse. Based on the observations of these tangent points, the prediction for the whole front is conducted. Hence, the apex of the CME is, if at all, only observed for a short period of time. In order to get an estimate of the CME Earth arrival we have to assume a designated shape of the CME front, which is in our case, an ellipse. As shown by \cite{Owens2017Nat} this assumption might not be valid since the CME interacts with the ambient solar wind.

\subsection{Effect of the ambient solar wind} \label{sec:ambientSW}
When considering different points along the idealized elliptical CME front, it is noteworthy that the ambient solar wind speeds at these points would likely be different. Furthermore, the part of the CME front corresponding to the greatest elongation as seen by STA and STB (i.e.~the points corresponding to the tangent to the CME front) would propagate in different ambient solar wind conditions. In Figure~\ref{fig:STASTBElon}, the modeled time-elongation profiles of the tangent points seen from STA (top panel) and STB (bottom panel) for event $\#$4 are shown. 

These profiles are obtained from one modeled ensemble member of the ELEvoHI prediction, separately for STA and STB (see Figure \ref{fig:MovieSnapshot}), and are therefore available from 2010-04-08 11:00 until 2010-04-14 10:00. As long as the CME front could be tracked in HI images (until about 2010-04-10 01:00), the plotted profiles are consistent with the measured HI time-elongation profiles, obtained using the SATPLOT tool. The colors represent the speed of the ambient solar wind at the corresponding points. Due to the propagation of the modeled CME in the heliosphere, the elongation of the tangent point ranges from roughly 8\textdegree~to about 92\textdegree~and the speed of the ambient solar wind at these points ranges from 330~km~s$^{-1}$ to 500~km~s$^{-1}$, with a maximum speed of 530~km~s$^{-1}$ at about 66\textdegree, for STA (top panel in Figure~\ref{fig:STASTBElon}). The range of the elongation is similar for STB (6\textdegree~to 91\textdegree) but the ambient solar wind speed ranges only from $\approx$450~km~s$^{-1}$ to $\approx$365~km~s$^{-1}$.

\begin{figure}[htp]

\centering
\includegraphics[width=1.0\textwidth]{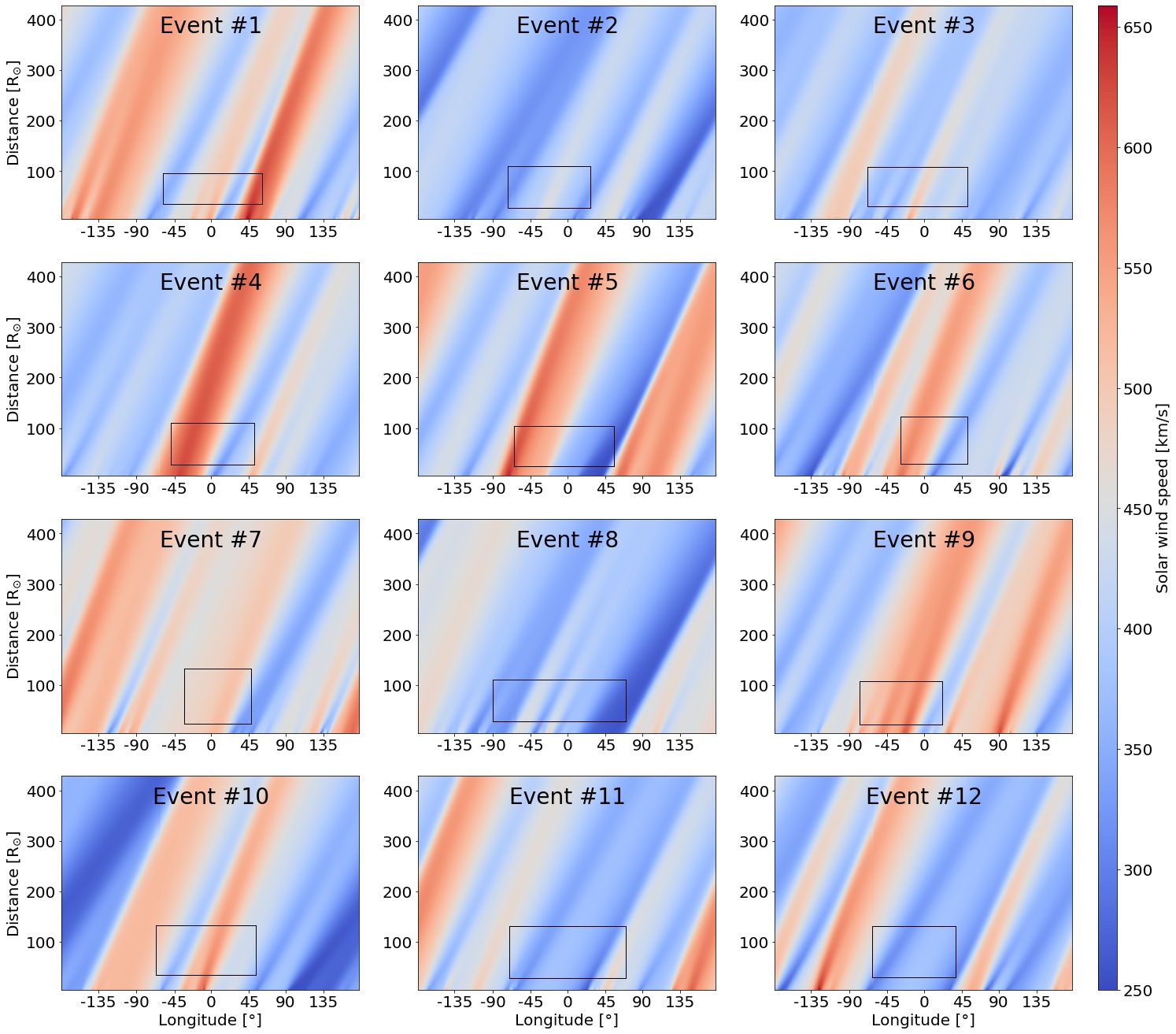}\hfill

\caption{Ambient solar wind speed provided by the WSA/HUX model combination for all 12 events under study. The black boxes define the areas that are used to estimate how structured the ambient solar wind is for each CME. Longitude of 0\textdegree\ corresponds to the longitude of Earth.}
\label{fig:AllSolarWinds}

\end{figure}

\begin{figure}[htp]

\centering
\includegraphics[width=1.0\textwidth]{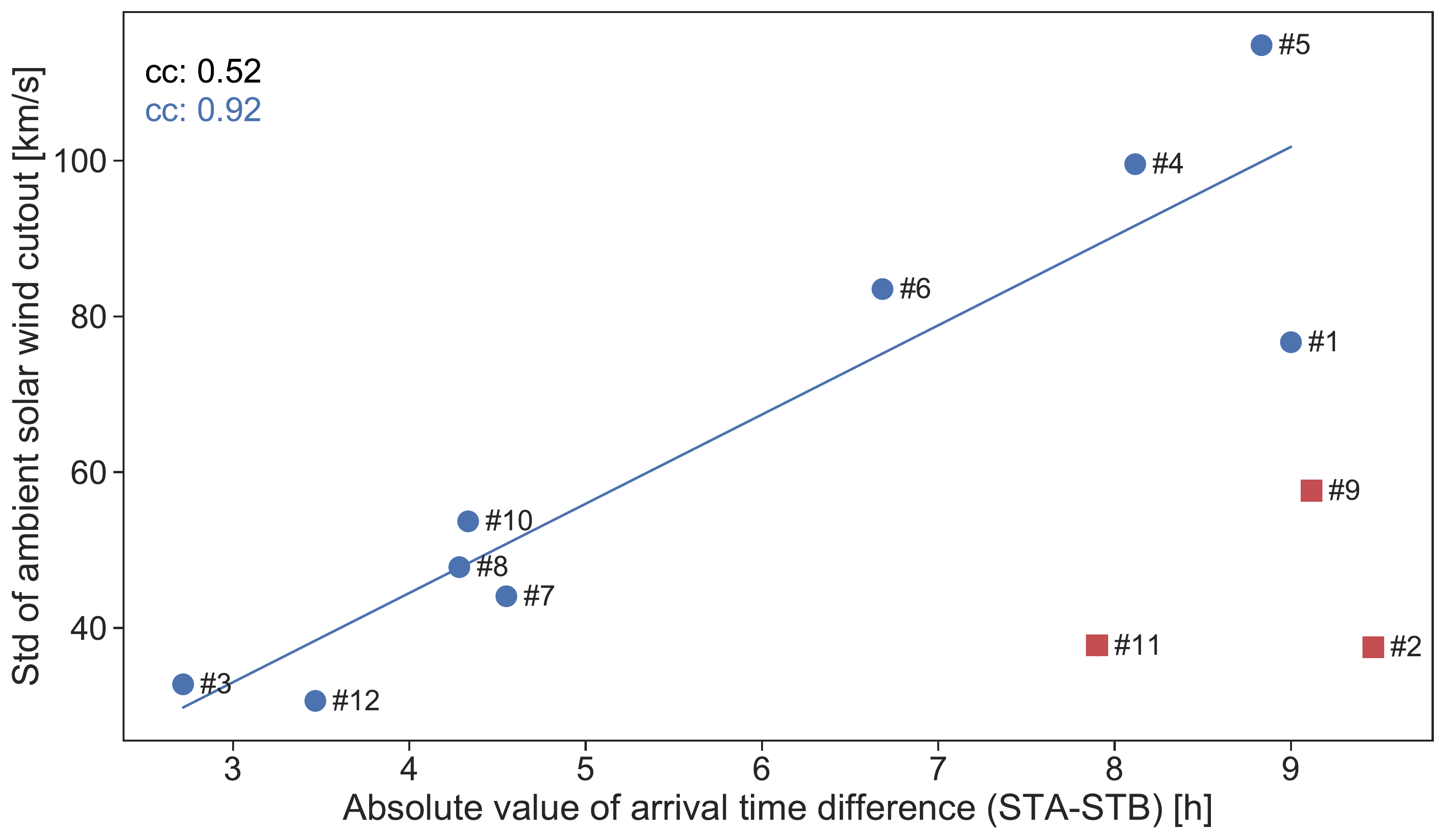}\hfill

\caption{Standard deviation of the ambient solar wind vs. the arrival time difference between STA and STB predictions. The Pearson correlation coefficient for all events under study (black) is calculated. In blue we present the Pearson correlation coefficient and a linear fit when excluding the outliers (indicated by the red boxes), i.e.~flank hits (events $\#2$ and $\#9$) and the CME-CME interaction event (event $\#11$).}
\label{fig:BGSW_ArrDiff}

\end{figure}
In the previous paragraph, we considered the ambient solar wind speed at the tangent point for one ensemble member. Additionally, we examine the distribution of the ambient solar wind speed considered for all ensemble members (see black boxes in Figures \ref{fig:WSA_BGSW} and \ref{fig:AllSolarWinds}) that are used as input to ELEvoHI for a single CME. From the areas framed by the black boxes, we calculate the standard deviation and correlate those to the absolute values of the difference between STA and STB arrival time predictions for each event (see Figure~\ref{fig:BGSW_ArrDiff}). This gives us the possibility to check the influence of the ambient solar wind on the arrival time differences. 
We obtain a Pearson correlation coefficient of $cc=0.52$ for all events under study. However, when excluding events $\#2$ and $\#9$, which are considered as "flank hits", and excluding event $\#11$ (CME-CME interaction event), the Pearson correlation coefficient increases to $cc=0.92$. This indicates that a more structured ambient solar wind (i.e.~a larger standard deviation) leads to a larger differences between STA and STB arrival time prediction.

\subsection{Comparison to in situ arrivals}
\begin{figure}[htbp]
\centering
\includegraphics[width=1.0\linewidth]{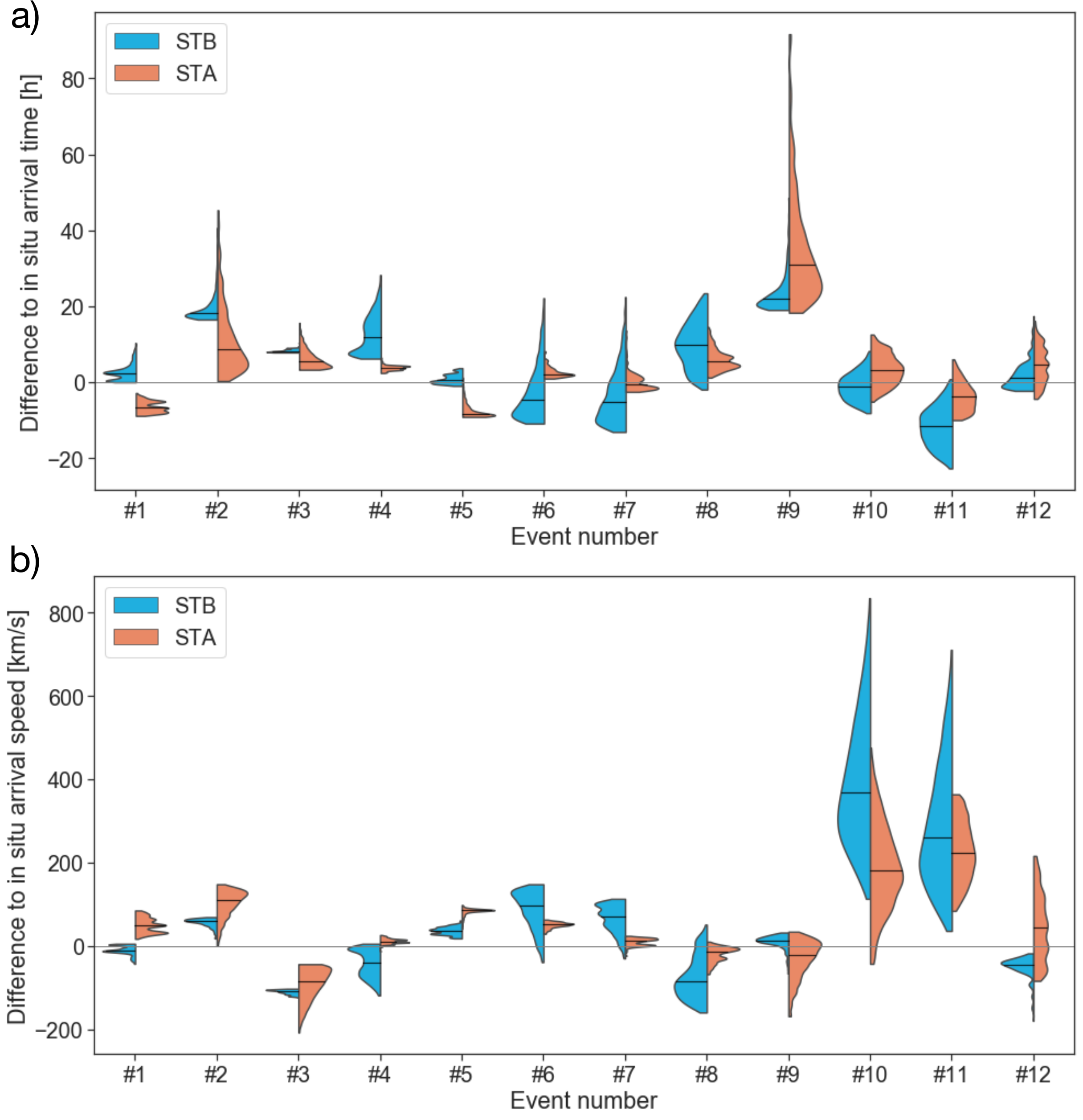}
\caption{Frequency distributions derived from all ensemble members for the arrival time prediction (top panel) and the arrival speed prediction (bottom panel) based on HI data from STB (blue) and STA (orange), respectively.
In the top panel, positive values correspond to a late arrival time prediction while negative values indicate an early arrival prediction. Positive/negative values in the bottom panel indicate an over-/underestimated arrival speed prediction. The black horizontal bars show the median values of the distributions of all the ensemble members for STB and STA.} \label{fig:ArrTimeViolin}
\end{figure}

Figure~\ref{fig:ArrTimeViolin} shows the distributions of the arrival time and arrival speed differences with respect to the in situ arrivals for all ensemble members for each CME. Blue and orange correspond to STB and STA ensemble predictions, respectively. The black horizontal lines indicate the median values of each distribution. When comparing the median predicted arrival times to the in situ arrivals, we obtain a mean absolute error (MAE) over all events of $7.5~\pm~9.5$~hrs and a root mean square error (RMSE) of $\approx10.4$~hrs. A mean error (ME) of $\approx4$~hrs indicates, in this setup, that ELEvoHI tends to predict the arrivals too late. The highest arrival time discrepancy is found for event $\#$9 where the prediction based on STA is 31~hrs too late. When comparing the median predicted arrival speeds to the in situ speeds we get a MAE of 87~$\pm$~111~km~s$^{-1}$, a RMSE of $\approx$123~km~s$^{-1}$ and a ME of $\approx$52~km~s$^{-1}$. The highest speed difference is found for the STB prediction of event $\#$10, overestimating the arrival speed by 369~km~s$^{-1}$.

Interestingly, event $\#$10 gives an accurate predicted arrival time, even though the predicted arrival speed is highly overestimated. When performing GCS reconstruction we obtain a high latitude and a large tilt angle for this CME meaning that the 3D propagation direction differs from that in the ecliptic plane (see Table \ref{tab:GCS}). As already mentioned, event $\#$11 is a CME-CME interaction event which explains the large discrepancy especially for the predicted arrival speed. The reason might be found in an extremely low drag due to preconditioning in the interplanetary space \citep{Rollett2014,Liu2014,Temmer2015}.

\section{Discussion and Conclusions} \label{Sec:Summary}

We present the ELEvoHI ensemble modeling results for 12 CMEs, occurring between February 2010 and July 2012, that were observed by both STEREO spacecraft. This study mainly focuses on the difference of the modeled arrival time and arrival speed when using STA and STB HI observations, separately. We find on average a difference of 6.5~hrs between arrival time predictions from the two spacecraft but the largest difference is about 9.5~hrs for event $\#9$. For the arrival speed we find a mean difference between STA and STB predictions of 63~km~s$^{-1}$ with a maximum difference of 189~km~s$^{-1}$ for event $\#10$. 

ELEvoHI tends to predict the arrival time later than observed for CMEs that are considered as 'flank hits' (event $\#2$ and event $\#9$). For such events the propagation direction with respect to Earth is larger than 20\textdegree, and not all of the ensemble members predict an Earth impact. The reason for the late arrival prediction may be found in the assumed circular shape (for $f=1.0$) and the highly curved flanks.

We provide two CME arrival time and arrival speed predictions, from STA and STB observation, for the same CME to examine the reasons for the discrepancy between these two predictions. We find, that the CME front propagates in different ambient solar wind conditions when observed in STA and STB HI images. However the kinematics of the CME front obtained e.g.~by STA data is used for modeling of the whole CME front, including the Earth-directed part. The same applies for predictions based on STB data, which is the reason for the differences in the predicted arrival times based on STA and STB observations.

We further see, that an ambient solar wind exhibiting a high variance within the area used for ELEvoHI model predictions leads to larger discrepancies between STA and STB model predictions. We obtain a Pearson correlation coefficient ($cc=0.92$), when excluding flank hits (events $\#2$ and event $\#9$) and the CME-CME interaction event (event $\#11$). Furthermore, we assume that in such cases the CME front is more likely to deform from an idealized elliptical shape due to interaction with the ambient solar wind \citep{Riley2004ApJ, Owens2017Nat}. 

The current CME forecasting abilities in the community are summarized in \cite{Riley2018}. They analyzed CME forecasts that have been submitted to the Community Coordinated Modeling Center (CCMC) \href{https://kauai.ccmc.gsfc.nasa.gov/CMEscoreboard/}{scoreboard} from 2013 to mid-2018. The CCMC scoreboard is a platform provided to scientists to compare their forecasts with each other in real-time. \cite{Riley2018} found that the CME shock arrival times for all models combined are predicted on average within $\pm10$~hrs but with standard deviations of sometimes more than 20~hrs. The best model performance was found for the WSA-ENLIL+Cone model \citep{Odstrcil2004}, run by the UK Met Office, having a bias of 1 hour, a MAE of 13~hrs and a standard deviation of 15~hrs. The results of this study are similar to the findings of \cite{Riley2018} when comparing the modeled arrival times to the actual arrivals of CMEs, as determined from in situ measurements. Here, we only perform hindcasts of CME arrivals. For the 24 arrival predictions (12 based on STA and 12 based on STB observations), we obtain a MAE of $7.5~\pm~9.5$~hrs, a RMSE of $\approx10.4$~hrs and a ME of $\approx4$~hrs for the arrival time. For the arrival speed, we get a MAE of 87~$\pm$~111~km~s$^{-1}$, a RMSE of $\approx$123~km~s$^{-1}$ and a ME of $\approx$52~km~s$^{-1}$. 

As already mentioned, event $\#$11 is a CME-CME interaction event studied e.g.~by \cite{Kubicka2016}. This CME was closely preceded by two other CMEs that erupted one and two days before this event and that altered the conditions in the heliosphere. The arrival time prediction for this CME is about 11~hrs too early, while the arrival speed is greatly overestimated (by 260~km~s$^{-1}$) using the ambient solar wind solutions provided by the WSA/HUX model. However, this model does not consider preceding CMEs and is likely not valid in such cases. An additional approach to infer the ambient solar wind conditions in the low heliosphere is shown in \cite{Barnard2019}. In this study the authors established a statistical relationship between the solar wind speed in the low heliosphere and the variability in HI images. A recent study by \cite{Amerstorfer2021} focuses on different input parameters to ELEvoHI including three possible methods to infer the ambient solar wind conditions needed by the model. First, the ambient solar wind speed is obtained from in-situ measurements at 1~AU. Second, the solar wind speed is based on an statistical approach using 14~years of OMNI data. Third, an estimate of the ambient solar wind speed is obtained from the WSA-HUX model combination. In their study, \cite{Amerstorfer2021} concluded that the third approach provides the best results. 

ELEvoHI provides ensemble predictions based on various inputs, namely propagation direction, half width, inverse aspect ratio and ambient solar wind speed. In the current version, ELEvoHI is not able to react to possible deflections of a CME during its propagation. Furthermore, the elliptical CME shape, once defined by the input parameters, does not change during propagation. This has been shown to be invalid by, for example, \cite{Rollett2014}, who performed a case study by combining HI data with in situ data to ascertain the kinematics of the 2012, March 7 CME. The authors demonstrated evidence for an asymmetric evolution of the CME, which was caused by the preconditioned ambient solar wind resulting in a different drag regime influencing different parts of the CME.

CME deflection \citep[e.g.][]{Wang2004,Wang2014} and deformation \citep[e.g.][]{Barnard2017,KayNievesChinchilla2020} are important factors when considering CME propagation in the heliosphere. The authors found that the failure to take these factors into account would likely lead to uncertainties in the arrival time and arrival speed prediction. \cite{Barnard2017} additionally showed that different tracks lead to quite different CME arrival time predictions. By using HI observations with better solar wind modeling and varying CME frontal shapes we should be able to improve our current arrival time predictions \citep{Barnard2020}.

A number of studies have taken advantage of stereoscopic HI observations, from the two STEREO spacecraft, to glean information on CME propagation and evolution \citep[e.g.][]{Liu2010A,Lugaz2010,Davies2013,Volpes2015}. 
\cite{Braga2020} studied 14 CMEs using the drag model and a modified version of the ELCon model, to get the CME parameters (e.g.~$\phi$, $\lambda$, $f$) based on HI observations from both STEREO spacecraft simultaneously. For the five events, that are also included in our study, we obtain similar results as \cite{Braga2020}. Please note that we run ELEvoHI in ensemble mode in this study.
We believe, that a stereoscopic view on CMEs incorporated in ELEvoHI will improve the arrival time predictions substantially.
Therefore, we strongly support ESA's L5 mission, equipped with a heliospheric imager \citep{Lavraud2016, Kraft2017}, and an additional heliospheric imager at L1. Fortunately, the upcoming Earth-orbiting PUNCH mission (launch planned in 2023) will also possess wide-angle white-light heliospheric imagers, as well as a coronagraph, and will be able to provide additional observations of CMEs. Based on information from these additional vantage points, more accurate CME arrival predictions are likely to be achieved. Since ELEvoHI is ready to be used in near real-time, future HI observations are essential for further CME arrival predictions. STA, currently near L5, will have moved beyond L4 by 2027, so it will be necessary to have heliospheric imagers that are observing the space between Sun and Earth after around 2030. 

In a next step, we want to further develop ELEvoHI in such a way that it can combine HI data from two vantage points in order to constrain the CME and exclude ensemble runs that are not consistent with the observations. Also the CME shape can be constrained by multiple HI observations and therefore, we aim to make the CME front deformable during the propagation through the heliosphere. Hence, the assumed elliptical CME front would be able to adjust according to the ambient solar wind conditions.
It was already shown in previous studies \citep[e.g.][]{Scott2019,Chi2020} that ghost fronts in the HI observations can be used to infer the structure of a CME. Using their approach, we also aim to improve our model by verifying and constraining the CME shape.

\section{Data Sources} \label{sec:DataSources}

\noindent \textbf{Data} \\

\noindent STEREO/HI: \url{https://www.ukssdc.ac.uk/solar/stereo/data.html} \\
\noindent STEREO/COR2 and SoHO/LASCO: \url{https://sdac.virtualsolar.org/cgi/} \\
\noindent NSO/GONG: \url{https://gong.nso.edu/data/magmap/} \\
\noindent HELCATS: \url{https://www.helcats-fp7.eu}\\
\noindent ICMECAT: \url{https://doi.org/10.6084/m9.figshare.6356420}

\noindent \textbf{Model} \\

\noindent ELEvoHI is available at \url{https://zenodo.org/record/3873420}. \\
\noindent EAGEL is available at \url{https://zenodo.org/record/4154458}.

\noindent \textbf{Results} \\

\noindent The visualization of each prediction result, i.e.\ movies and figures, can be downloaded from \url{https://doi.org/10.6084/m9.figshare.12758312}.

\noindent \textbf{Software} \\

\noindent IDL\textsuperscript{TM} Version 8.4 \\
\noindent Python 3.7.6 \\
\noindent SATPLOT: \url{https://hesperia.gsfc.nasa.gov/ssw/stereo/secchi/idl/jpl/satplot/SATPLOT\_User\_Guide.pdf}

%
%
%
%
%
%
%
%

\acknowledgments
T.A., J.H., M.B., M.A.R., C.M., A.J.W., R.L.B, and U.V.A. thank the Austrian Science Fund (FWF): P31265-N27, J4160-N27, P31659-N27, P31521-N27.


%
\bibliography{juh_bib}




\end{document}